\documentclass[a4paper,12pt]{article}
\usepackage{amssymb,amsmath}
\usepackage[dvips]{lscape,graphicx}

\newcommand{\bc}{\begin{center}}
\newcommand{\ec}{\end{center}}
\newcommand{\bd}{\begin{displaymath}}
\newcommand{\ed}{\end{displaymath}}
\newcommand{\be}{\begin{equation}}
\newcommand{\ee}{\end{equation}}
\newcommand{\ba}{\begin{array}}
\newcommand{\ea}{\end{array}}
\newcommand{\bt}{\begin{tabular}}
\newcommand{\et}{\end{tabular}}

\newcommand{\ds}{\displaystyle}

\begin{document}

\bibliographystyle{OurBibTeX}

\begin{titlepage}

\vspace*{-15mm}
\begin{flushright}
UH511-1164-2011\\
\end{flushright}
\vspace*{5mm}

\begin{center}
{\sffamily \Large
Dark Energy density in models with Split Supersymmetry and degenerate vacua}
\\[6mm]
C.~Froggatt${}^{a}$,
R.~Nevzorov${}^{b}$\footnote{On leave of absence from the Theory Department, ITEP, Moscow, Russia.},
H.~B.~Nielsen${}^{c}$\\[3mm]
{\small\it
${}^{a}$ School of Physics and Astronomy, University of Glasgow,
Glasgow, UK\\[2mm]
${}^{b}$ Department of Physics and Astronomy, University of Hawaii,
Honolulu, USA\\[2mm]
${}^{c}$ The Niels Bohr Institute, University of Copenhagen, 
Copenhagen, Denmark
}\\[1mm]
\end{center}
\vspace*{0.75cm}



\begin{abstract}{
\noindent
In $N=1$ supergravity supersymmetric (SUSY) and non-supersymmetric
Minkowski vacua originating in the hidden sector can be degenerate.
In the supersymmetric phase in flat Minkowski space non-perturbative
supersymmetry breakdown may take place in the observable sector,
inducing a non-zero and positive vacuum energy density. Assuming that
such a supersymmetric phase and the phase in which we live are degenerate,
we estimate the value of the cosmological constant. We argue that the
observed value of the dark energy density can be reproduced in the
Split-SUSY scenario of the supersymmetry breaking if the SUSY breaking
scale is of order of $10^{10}\,\mbox{GeV}$.
}
\end{abstract}

\end{titlepage}

\newpage
\section{Introduction}

The tiny value of the cosmological constant (dark energy), which is responsible
for the accelerated expansion of the Universe and constitutes $70\%-73\%$
of its energy density, is a major puzzle for modern particle physics.
A fit to the recent data shows that its value is
$\Lambda \sim 10^{-123}M_{Pl}^4 \sim 10^{-55} M_Z^4$ \cite{1}.
At the same time much larger contributions should come
from electroweak symmetry breaking ($\sim 10^{-67}M_{Pl}^4$)
and QCD condensates ($\sim 10^{-79}M_{Pl}^4$). Moreover the
contribution of zero--modes is expected to push the
vacuum energy density even higher up to $\sim M_{Pl}^4$, i.e.
\begin{equation}
\rho_{\Lambda}\simeq \sum_{bosons}\frac{\omega_b}{2}- \sum_{fermions}\frac{\omega_f}{2}=
\int_{0}^{\Omega}\biggl[\sum_b\sqrt{|\vec{k}|^2+m_b^2}
-\sum_f\sqrt{|\vec{k}|^2+m_f^2}\biggr]\frac{d^3\vec{k}}{2(2\pi)^3}\sim -\Omega^4\,,
\label{01}
\end{equation}
where the $m_b$ and $m_f$ are the masses of bosons and fermions while
$\Omega\sim M_{Pl}$. Because of the enormous cancellation needed between
the contributions of different condensates to $\Lambda$, the smallness of
the cosmological constant should be regarded as a fine--tuning problem.

The smallness of the cosmological constant could be related to an almost
exact symmetry. An exact global supersymmetry (SUSY) ensures zero value
for the energy density at the minimum of the potential of the scalar fields.
However the breakdown of supersymmetry induces a huge and positive
contribution to the total vacuum energy density of order $M_{S}^4$, where
$M_{S}$ is the SUSY breaking scale. Because superpartners of quarks and
leptons have not yet been observed, $M_{S}$ is expected to be higher
than $100\,\mbox{GeV}$.

Since we are interested in the value of the cosmological constant, we must
include gravity and thus local supersymmetry in our consideration.
In supergravity (SUGRA) an enormous fine--tuning is also required to keep
the cosmological constant around its observed value \cite{2}. Our basic
scenario for evaluating the tiny value of the cosmological constant implies
that the SUGRA scalar potential of the hidden sector has a supersymmetric
Minkowski minimum (second vacuum), in addition to the physical vacuum in
which we live. In this second vacuum the supersymmetry is expected to be
broken dynamically in the observable sector when the supersymmetric QCD
interaction becomes non-perturbative. The physical and the second vacua
are assumed to be degenerate, so that the estimated vacuum energy density
in the second vacuum is then transferred to our vacuum.

The assumed degeneracy of the vacua is supposed to be justified by the
so-called Multiple Point Principle (MPP). According to the MPP, Nature
chooses values of coupling constants such that many phases of the
underlying theory should coexist \cite{10}. On the phase diagram of the
theory it corresponds to a special point -- the multiple point.
The vacuum energy densities of these different phases are degenerate
at the multiple point. In this paper we shall apply MPP to $(N=1)$
supergravity to give degeneracy between the two vacua mentioned above.
This would normally require an extra fine-tuning associated with the
presence of the second vacuum \cite{Froggatt:2003jm}. However, the breakdown
of global symmetries, which are chosen to protect a zero value for the
cosmological constant in SUGRA models, may lead to a set of degenerate
vacua with broken and unbroken supersymmetry (SUSY) whose vacuum energy
densities vanish in the leading approximation
\cite{Froggatt:2004gc}-\cite{Froggatt:2005nb}, resulting in a natural
realization of the MPP conditions \cite{10}. In our previous articles
the MPP assumption was used to estimate the value of the cosmological
constant in such SUGRA models inspired by degenerate vacua
\cite{Froggatt:2003jm}-\cite{talks}. It is the main purpose of the
present article to use the same idea for estimating the cosmological
constant, but now under the assumption of {\em a Split SUSY picture}.

Instead of just postulating the MPP, a large set of approximately degenerate
vacua with broken and unbroken supersymmetry can also arise from the vast
landscape of string theory vacua, if the vacua with large cosmological
constants are not allowed. Recently it was realized that the presence of an
enormous number of long-lived metastable vacua favors high-scale breaking
of supersymmetry, which motivated the introduction of the Split SUSY
scenario. In this paper we attempt to estimate the value of the dark energy
in the Split SUSY model, assuming the  degeneracy of vacua  with
broken and unbroken supersymmetry (i.e. MPP). We argue that the observed value
of the cosmological constant can be reproduced in the considered case,
if the SUSY breaking scale is $M_S\sim 10^{9}-10^{10}\,\mbox{GeV}$. This
Split SUSY scenario predicts extremely long-lived gluinos.

The paper is organized as follows. In the next section we present $(N=1)$
SUGRA models in which the MPP conditions are fulfilled. In section 3 we discuss
the Split SUSY scenario, and in section 4 we present some numerical estimates
of the vacuum energy density and consider possible phenomenological
implications. Our results are summarized in section 5.

\section{SUGRA models inspired by degenerate vacua}
\label{noscale}

The full $(N=1)$ SUGRA Lagrangian \cite{4}-\cite{3} is specified
in terms of an analytic gauge kinetic function $f_a(\phi_{M})$ and
a real gauge-invariant K$\Ddot{a}$hler function
$G(\phi_{M},\phi_{M}^{*})$, which depend on the chiral superfields
$\phi_M$. The function $f_{a}(\phi_M)$ determines the gauge
coupling constants $Re f_a(\phi_M)=1/g_a^2$, where the index $a$
designates different gauge groups. The K$\Ddot{a}$hler function is
a combination of two functions
\be
G(\phi_{M},\phi_{M}^{*})=K(\phi_{M},\phi_{M}^{*})+\ln|W(\phi_M)|^2\,,
\label{1}
\ee
where $K(\phi_{M},\phi_{M}^{*})$ is the K$\Ddot{a}$hler potential
while $W(\phi_M)$ is the superpotential of the considered SUSY model.
Here we use standard supergravity mass units: $\ds\frac{M_{Pl}}{\sqrt{8\pi}}=1$.

The SUGRA scalar potential is given by
\begin{equation}
\begin{array}{c}
V(\phi_M,\phi^{*}_M)=\sum_{M,\,\bar{N}} e^{G}\left(G_{M}G^{M\bar{N}}
G_{\bar{N}}-3\right)+\ds\frac{1}{2}\sum_{a}(D^{a})^2,\,\,\,
D^{a}=g_{a}\sum_{i,\,j}\left(G_i T^a_{ij}\phi_j\right)\,,\\[2mm]
G_M \equiv \ds\frac{\partial G}{\partial \phi_M}\,,\quad
G_{\bar{M}}\equiv \ds\frac{\partial G}{\partial \phi^{*}_M}\,,\quad
G_{\bar{N}M}\equiv \ds\frac{\partial^2 G}{\partial \phi^{*}_N \partial \phi_M}\,,\quad
G^{M\bar{N}}=G_{\bar{N}M}^{-1}\,,
\end{array}
\label{2}
\end{equation}
where $g_a$ is the gauge coupling constant associated with the
generator $T^a$ of the gauge transformations. In order to break
supersymmetry in $(N=1)$ SUGRA models, a hidden sector is
introduced. It is assumed that the superfields of the hidden
sector $(z_i)$ interact with the observable ones only by means of
gravity. At the minimum of the scalar potential (\ref{2}), hidden
sector fields acquire vacuum expectation values (VEVs) so that at least
one of their auxiliary fields
\be
F^{M}=e^{G/2}G^{M\bar{P}}G_{\bar{P}}
\label{201}
\ee
gets a non-vanishing VEV, giving rise to the breakdown of local SUSY and
generating a non--zero gravitino mass $m_{3/2}=<e^{G/2}>$.

As mentioned in section 1, the successful implementation of the MPP in
$(N=1)$ supergravity requires us to assume the existence of a vacuum in
which the low--energy limit of the considered theory is described by
a pure supersymmetric model in flat Minkowski space. According to the
MPP this vacuum and the physical one must be degenerate. Such a second
vacuum is realised only if the SUGRA scalar potential has a minimum
where $m_{3/2}=0$. The corresponding minimum is achieved when
the superpotential $W$ for the hidden sector and its derivatives
vanish, i.e.
\be
W(z_m^{(2)})=0\,,
\label{202}
\ee
\be
\frac{\partial W(z_i)}{\partial z_m}\Biggl|_{z_m=z_m^{(2)}}=0
\label{203}
\ee
where $z_m^{(2)}$ denote vacuum expectation values of the hidden sector
fields in the second vacuum. In general Eq.~(\ref{202}) represents
the extra fine-tuning associated with the presence of the
supersymmetric Minkowski vacuum.

The simplest K$\Ddot{a}$hler potential and superpotential
that satisfies conditions (\ref{202}) and (\ref{203})
can be written as
\be
K(z,\,z^{*})=|z|^2\,,\qquad\qquad W(z)=m_0(z+\beta)^2\,.
\label{204}
\ee
The hidden sector of this SUGRA model contains only one singlet
superfield $z$. If the parameter $\beta=\beta_0=-\sqrt{3}+2\sqrt{2}$,
the SUGRA scalar potential of the considered model possesses
two degenerate minima with zero energy density at the classical
level. One of them is a supersymmetric Minkowski minimum that
corresponds to $z^{(2)}=-\beta$. In the other minimum of the
SUGRA scalar potential ($z^{(1)}=\sqrt{3}-\sqrt{2}$)
local supersymmetry is broken; so it can be associated
with the physical vacuum. Varying the parameter $\beta$ around
$\beta_0$ one can obtain a positive or a negative contribution
from the hidden sector to the total energy density of the physical
vacuum. Thus $\beta$ can be fine--tuned so that the
physical and second vacua are degenerate.

The obvious drawback of the model discussed above is related
with the degree of the fine-tuning which is required, in order to
arrange for the presence of a supersymmetric Minkowski vacuum
as well as for the degeneracy of the physical and second vacua.
Nevertheless in some SUGRA models this fine-tuning can be
alleviated. Let us consider the no--scale inspired SUGRA model
with two hidden sector fields ($T$ and $z$) and a set of
chiral supermultiplets $\varphi_{\sigma}$ in the observable sector.
These superfields transform differently under the imaginary translations
($T\to T+i\beta,\, z\to z,\, \varphi_{\sigma}\to \varphi_{\sigma}$)
and dilatations
($T\to\alpha^2 T,\, z\to\alpha\,z,\,\varphi_{\sigma}\to\alpha\,\varphi_{\sigma}$),
which are subgroups of the $SU(1,1)$ group \cite{15}--\cite{12}
\footnote{The Lagrangian for a no--scale SUGRA model is invariant
under imaginary translations and dilatations. As a consequence the
vacuum energy density goes to zero near global minima of the
tree--level scalar potential which vanishes identically along
some directions \cite{4}, \cite{12}--\cite{11}. Thus imaginary
translations and dilatations protect a zero value for the
cosmological constant in supergravity. However these symmetries
also preserve supersymmetry in all vacua, which has to be broken
in any phenomenologically acceptable theory.}.
In order to ensure the appropriate breakdown of local supersymmetry,
we assume that there is a weak breaking of the dilatation invariance
of the hidden sector superpotential characterised by an hierarchically
small parameter $\varkappa$. The full superpotential of the model
is given by \cite{Froggatt:2005nb}:
\be
\ba{c}
W(z,\,\varphi_{\alpha})=W_{hid} + W_{obs}\,,\\[2mm]
W_{hid}=\varkappa\biggl(z^3+\mu_0 z^2+\sum_{n=4}^{\infty}c_n z^n\biggr),\qquad
W_{obs}=\ds\sum_{\sigma,\beta,\gamma}\ds\frac{1}{6}
Y_{\sigma\beta\gamma}\varphi_{\sigma}\varphi_{\beta}\varphi_{\gamma}\,.
\ea
\label{191}
\ee
The superpotential (\ref{191}) contains a bilinear mass term for the
superfield $z$ and higher order terms $c_n z^n$ that spoil dilatation
invariance. A term proportional to $z$ is not included. It can be
forbidden by a gauge symmetry of the hidden sector, if $z$ transforms
non--trivially under the corresponding gauge transformations. Here we
do not allow the breakdown of dilatation invariance in the superpotential
of the observable sector, in order to avoid the appearance of potentially
dangerous terms which lead, for instance, to the so--called $\mu$--problem.

We also assume that the dilatation invariance is broken in the K$\ddot{a}$hler
potential of the observable sector, so that the full K$\ddot{a}$hler
potential takes the form \cite{Froggatt:2005nb}:
\be
\ba{rcl}
K(\phi_{M},\phi_{M}^{*})&=&\ds-3\ln\biggl[T+\overline{T}
-|z|^2-\sum_{\alpha}\zeta_{\alpha}|\varphi_{\alpha}|^2\biggr]
+\\[2mm]
&+&\sum_{\alpha, \beta}\biggl(\ds\frac{\eta_{\alpha\beta}}
{2}\,\varphi_{\alpha}\,\varphi_{\beta}+h.c.\biggr)+
\sum_{\beta}\xi_{\beta}|\varphi_{\beta}|^2\,,
\ea
\label{192}
\ee
where $\zeta_{\alpha}$, $\eta_{\alpha\beta}$, $\xi_{\beta}$ are some
constants. In the limit when $\eta_{\alpha\beta}$, $\xi_{\beta}$
and $\varkappa$ go to zero, the dilatation invariance is restored,
protecting supersymmetry and a zero value of the cosmological constant.
It is worth noticing that we only allow the breakdown of the dilatation
invariance in the K$\ddot{a}$hler potential of the observable sector,
since any variations in the K$\ddot{a}$hler potential of the hidden
sector may spoil the vanishing of the vacuum energy density in global
minima. We restrict our consideration to the simplest set of terms
that break dilatation invariance in the K$\ddot{a}$hler potential.
Additional terms which are proportional to $|\varphi_{\alpha}|^2$
normally appear in minimal SUGRA models \cite{minimal-sugra}. The
other terms $\eta_{\alpha\beta}\varphi_{\alpha} \varphi_{\beta}$
introduced in the K$\Ddot{a}$hler potential (\ref{192}) give rise
to effective $\mu$ terms after the spontaneous breakdown of local
supersymmetry, solving the $\mu$ problem \cite{30}.

In the considered SUGRA model the scalar potential is positive definite
\be
V=V_{hid}+V_{obs},
\label{1921}
\ee
\be
V_{hid}=\frac{1}{3(T+\overline{T}-|z|^2)^2}
\biggl|\frac{\partial W_{hid}(z)}{\partial z}\biggr|^2,
\label{1922}
\ee
\be
V_{obs}=\sum_{\alpha}\biggl|\ds\frac{\partial
W_{eff}(y_{\beta})}{\partial y_{\alpha}}
+m_{\alpha}y^{*}_{\alpha}\biggr|^2+\ds\frac{1}{2}\sum_{a}(D^{a})^2\,,
\label{193}
\ee
so that the vacuum energy density vanishes near its global minima.
In Eq.~(\ref{193}) $y_{\alpha}$ are canonically normalized scalar fields
\be
\ba{c}
y_{\alpha}=\tilde{C}_{\alpha}\varphi_{\alpha}\,,\qquad
\tilde{C}_{\alpha}=\xi_{\alpha}\biggl(1+\ds\frac{1}
{x_{\alpha}}\biggr)\,,\qquad x_{\alpha}
=\frac{\xi_{\alpha}<(T+\overline{T}-|z|^2)>}{3\zeta_{\alpha}}\,.
\ea
\label{196}
\ee
The mass parameters $m_{\alpha}$ are given by \cite{Froggatt:2005nb}
\be
m_{\alpha} = \frac{m_{3/2} x_{\alpha}}{1+x_{\alpha}},
\ee
where $m_{3/2}$ denotes the gravitino mass.
The effective superpotential, which describes the interactions of
observable superfields at low energies, is given by
\be
\ba{c}
W_{eff}=\sum_{\alpha,\,\beta}\ds\frac{\mu_{\alpha\beta}}{2}\,
y_{\alpha}\,y_{\beta}+\sum_{\alpha,\,\beta,\,\gamma}
\ds\frac{h_{\alpha\beta\gamma}}{6}\,y_{\alpha}\,y_{\beta}\,y_{\gamma}\,,\\[3mm]
\mu_{\alpha\beta}=m_{3/2}\eta_{\alpha\beta}(\tilde{C}_{\alpha}
\tilde{C}_{\beta})^{-1}\,,\qquad\qquad
h_{\alpha\beta\gamma}=\ds\frac{Y_{\alpha\beta\gamma}
(\tilde{C}_{\alpha}\tilde{C}_{\beta}\tilde{C}_{\gamma})^{-1}}
{<(T+\overline{T}-|z|^2)^{3/2}>}\,.
\ea
\label{197}
\ee

The form of the superpotential (\ref{191}) guarantees that there is
always a supersymmetric Minkowski minimum at $z=0$, because the
conditions (\ref{202}) and (\ref{203}) are fulfilled for $z=0$
without any extra fine-tuning. In the simplest case when $c_n=0$,
$V_{hid}$ has two minima, at $z=0$ and $z=-\frac{2\mu_0}{3}$.
At these points the scalar potential (\ref{1922}) achieves its
absolute minimal value i.e.~zero. In the first vacuum, where $z=-\frac{2\mu_0}{3}$,
local supersymmetry is broken so that the gravitino becomes massive
\begin{equation}
m_{3/2}=\biggl<\frac{W(z)}{(T+\overline{T}-|z|^2)^{3/2}}\biggr>
=\frac{4\kappa\mu_0^3}{27\biggl<\biggl(T+\overline{T}
-\frac{4\mu_0^2}{9}\biggr)^{3/2}\biggr>}\,.
\label{7}
\end{equation}
and all scalar particles get non--zero masses
$m_{\sigma}\sim \frac{m_{3/2} \xi_{\sigma} }{\zeta_{\sigma}}$. Assuming
that $\xi_{\alpha}$, $\zeta_{\alpha}$, $\mu_0$ and $<T>$ are all of
order unity, a supersymmetry breaking scale $M_S\sim 1\,\mbox{TeV}$
can only be obtained for extremely small values of $\varkappa\simeq 10^{-15}$.
In the second minimum, with $z=0$, the superpotential of the hidden sector
vanishes and local SUSY remains intact, so that the low--energy limit of
this theory is described by a pure SUSY model in flat Minkowski space.
If the high order terms $c_n z^n$ are present in Eqs.~(\ref{191}),
the scalar potential of the hidden sector may have many degenerate vacua,
with broken and unbroken supersymmetry, in which the vacuum energy
density vanishes.

In the first vacuum, where the gravitino gains a non-zero mass $m_{3/2}$,
the breakdown of electroweak (EW) symmetry can also be achieved, if the
Higgs sector involves two Higgs doublets ($H_u$ and $H_d$) and a
singlet field $S$. The most general superpotential that describes
the interactions of the Higgs superfields at low energies can be
written as
\begin{equation}
W_{H}=\lambda S (H_u H_d) - \dfrac{\sigma}{3} S^3 - \mu (H_u H_d)
- \dfrac{\mu'}{2} S^2.
\label{8}
\end{equation}
We assume that the Lagrangian of the considered model is invariant under
permutation symmetry $H_u\leftrightarrow H_d$ so that the soft scalar
masses $m_{H_u}^2$ and $m_{H_d}^2$ are equal, i.e. $m_{H_u}=m_{H_d}=m_H$.
Then the scalar potential (\ref{193}) has a global minimum with zero
vacuum energy density at which the EW symmetry is broken by
the vacuum expectation values (VEVs) of the Higgs fields
\begin{equation}
<S>=s=\dfrac{\mu-m_H}{\lambda}\,,\qquad\quad
<H_u>=<H_d>=\pm\sqrt{\dfrac{\sigma s^2 + \mu' s - m_S s}{\lambda}}\,,
\label{9}
\end{equation}
where $m_S$ is the soft scalar mass of the singlet field $S$. The scalar
potential also has another minimum with the same energy density, where
the Higgs field VEVs all vanish and in which the EW symmetry remains intact.
In the supersymmetric vacuum $\mu$, $\mu'$, $m_H$ and $m_S$ vanish and
the Higgs fields do not acquire VEVs, so that EW symmetry does not get broken.
Thus we find that there are three degenerate vacua: the supersymmetric
Minkowski vacuum and two with SUSY broken in the hidden sector, one of
which has unbroken EW symmetry and the other has broken EW symmetry as
in the physical vacuum. As a consequence the MPP conditions are fulfilled
automatically without any extra fine-tuning at the tree--level.

It is worth emphasizing here that, even after the EW symmetry breaking,
the total vacuum energy density remains zero at the minimum where SUSY is
broken so that vacua with broken and unbroken supersymmetry are degenerate.
Thus, in the considered case, the breakdown of dilatation invariance provides
a mechanism that leads to the cancellation of the contribution to the
total vacuum energy density arising from the breakdown of the EW symmetry.
At the same time this breakdown of dilatation invariance also ensures the
cancellation of the vacuum energy density which originates from the breakdown
of local supersymmetry.

Of course, this remarkable cancellation takes place at the tree level only.
The inclusion of perturbative and non--perturbative corrections to the considered
SUGRA Lagrangian, which should depend on the structure of the underlying theory,
are expected to spoil the degeneracy of vacua inducing a huge energy density
in the vacuum where SUSY is broken. Moreover the considered SUGRA model has
other serious shortcomings as well. In particular, the mechanism for the
stabilization of the vacuum expectation value of the hidden sector field $T$
remains unclear. As a result the gravitino mass (see Eq.~(\ref{7})) and
the supersymmetry breaking scale are not fixed in the vacuum where SUSY is broken.
Therefore the SUGRA model discussed above should be considered as a toy example.
This model demonstrates that, in $(N=1)$ supergravity, there might be a mechanism
which ensures the cancellation of different contributions to the total vacuum
energy density in the physical vacuum. This mechanism may also lead to a set
of degenerate vacua with broken and unbroken supersymmetry,
resulting in a natural realization of the multiple point principle.

\section{Split SUSY and the Landscape}
\label{landscape}

In the present section we want to discuss the Split SUSY scenario,
which we shall use in section 4 for our estimate of the cosmological
constant.
For this estimation it is crucial that we assume that the
gauge couplings $1/g_a^2= Re f_a(T,\, z)$ are the same in the phase
in which we live and in the supersymmetric phase. This assumption is of course
only justified if the kinetic functions $f_a(T,\, z)$ are approximately
constant. Due to the
mild dependence of $f_a(T,\, z)$ on $T$ and $z$ the derivatives of the gauge
kinetic function tend to be small. As a result the gauginos are typically
substantially lighter than the scalar particles in the considered SUSY
models, i.e. $M_a\ll m_{\alpha}$. Such a hierarchical structure of the
particle spectrum naturally appears in models with Split Supersymmetry
\footnote{It is worth noting that it seems to be rather problematic to
generate a large mass hierarchy between the gauginos and sfermions within
the toy SUGRA models discussed in section 2. Indeed, even in the limit,
when $f_a(T,\, z)=const$, the non--zero gaugino masses are expected to get
induced \cite{AMSB}. Therefore, in our case an additional fine-tuning is
probably required to ensure that $\frac{M_a}{m_{\alpha}}\ll 10^{-2}$.}.

In Split SUSY one gives up on the naturalness criterion in dealing with
the electroweak scale \cite{ArkaniHamed:2004fb}-\cite{Giudice:2004tc}.
The corresponding idea has its origin in the observation that all known
models of electroweak symmetry breaking, including supersymmetric ones,
require an incredible amount of fine-tuning of the vacuum energy, such that
the resulting cosmological constant is as small as observed. This amount of
fine-tuning is much more severe than what is required to make the Higgs
boson mass free of quadratic divergence. This suggests the possibility that
our notions of naturalness might be misleading, and that other fine-tuning
mechanisms may be at work in nature.

In the Split SUSY scenario supersymmetry is not used to stabilize the weak
scale. This stabilization is supposed to be provided by some other
fine-tuning mechanism, which anyway is needed to explain the value of
the cosmological constant. Therefore, in the
considered models, the SUSY breaking scale $M_S$ is taken to be much above
10 TeV. All scalar particles acquire masses at this high scale $M_S$, except
for a single neutral Higgs boson, whose mass is either finely-tuned or is
preserved by some other mechanism. The gauginos and Higgsinos of this theory
are chosen to lie near the TeV scale, so as to ensure gauge coupling unification at
$M_{GUT}\sim 10^{16}\,\mbox{GeV}$. Indeed, if all sfermions are heavy, the
pattern of grand unification is unchanged as compared with the MSSM, since
heavy sfermions form complete $SU(5)$ representations \cite{Giudice:2004tc}.
Such a modification of the MSSM spectrum does not necessarily affect the mass
parameters of gauginos and Higgsinos, which can be protected by a combination
of an R-symmetry and Peccei Quinn symmetry, and thus they can have weak-scale masses.
Hence, a TeV-scale lightest neutralino can be an appropriate cold dark matter
candidate \cite{Giudice:2004tc}-\cite{dm-split-susy}. A Split SUSY spectrum
naturally arises in frameworks where the mechanism of SUSY breaking preserves
an R-symmetry and/or forbids Gaugino and Higgsino mass terms \cite{ArkaniHamed:2004fb},
\cite{ArkaniHamed:2004yi}, \cite{split-susy-spectrum}.

Thus the Split SUSY scenario retains the successes of the MSSM, aside from the
tuning of the light Higgs mass. At the same time some flaws inherent to the MSSM
elegantly disappear, when the scalar superpartners decouple. The ultra-heavy scalars,
whose masses are a priori undetermined and can in principle range from hundreds
of TeV up to $10^{13}\,\mbox{GeV}$ \cite{Giudice:2004tc}, guarantee the absence
of large supersymmetric flavor changing interactions and CP violation. The
generic constraints from flavour and electric dipole moment data, which set
$M_S>100-1000\,\mbox{TeV}$, are easily satisfied within this class of models.
The dimension-five operators which mediate proton decay are also suppressed,
delaying proton decay which now occurs via dimension-six operators.
The increase in the predicted proton lifetime is also in better
agreement with data \cite{Murayama:2001ur}.

Nevertheless, since the sfermions are ultra-heavy, the large quadratic corrections to
the mass of the Higgs are not cancelled in the manner present in weak-scale
supersymmetry. As a consequence, in Split SUSY the Higgs sector remains
extremely fine-tuned, often with the understanding that this fine-tuning
could be resolved by some anthropic-like selection effects \cite{Weinberg:1987dv}.
In other words the solution to both the hierarchy and cosmological constant
problems might not involve natural cancellations, but follow from a
completely different reasoning, such as the idea that galaxy and
star formation, chemistry and biology, are simply impossible without these
scales having the values found in our Universe
\cite{Weinberg:1987dv}-\cite{anthropic-principle}. In this case SUSY may have
nothing to do with the naturalness problem, although it may be a necessary
ingredient in a fundamental theory of nature such as String Theory.

It has been argued that String Theory can have an enormous number of long-lived metastable
vacua \cite{landscape}-\cite{Susskind:2004uv}. The space of such string theory vacua
is called the ``landscape". The number of discrete vacua in String Theory is measured
not in the millions or billions but in googles ($\sim 10^{100}$)
\cite{Bousso:2000xa}-\cite{Susskind:2004uv}. Recent developments in String Theory
applied a statistical approach to the large multitude of universes, corresponding to
the ``landscape" of vacua present in the theory
\cite{large-number-of-vacua}-\cite{high-energy-susy-breaking}.
These investigations indicate that among the vast number of metastable vacua, there
can be a small subset exhibiting low scale SUSY breaking. However the fine-tuning
required to achieve a small cosmological constant implies the need for a much larger
number of such vacua. Remarkably, the total number of vacua in String Theory can
be large enough to fine-tune both the cosmological constant and the Higgs mass,
favoring a high-scale breaking of supersymmetry
\cite{high-energy-susy-breaking}-\cite{Susskind:2004uv}. It is thus statistically
feasible in String Theory for us to live in a universe fine-tuned in the way we
find it, thereby having both a small cosmological constant and the electroweak scale
stabilized in the 100 GeV range. In the vast landscape of possible string theory vacua,
we may find ourselves in the observed ground state simply because of a cosmic
selection rule, i.e. the anthropic principle \cite{Weinberg:1987dv}.

The idea of the multiple point principle and the landscape paradigm have at least two
things in common. Both approaches imply the existence of a large number of vacua with
broken and unbroken supersymmetry. The landscape paradigm and MPP also imply that the
parameters of the theory, which leads to the Standard Model (SM) at low energies, can be extremely
fine-tuned so as to ensure a tiny vacuum energy density and a large hierarchy between $M_{Pl}$
and the electroweak scale. Moreover the MPP assumption may originate from the landscape of
string theory vacua, if all vacua with a large cosmological constant are forbidden
for some reason so that all the allowed string vacua, with broken and unbroken supersymmetry,
are approximately degenerate. If this is the case, then the breaking of supersymmetry at
high scales is probably still favored. Therefore our attempt to estimate the value of
the cosmological constant within the Split SUSY scenario assuming the degeneracy of
vacua with broken and unbroken supersymmetry might not be so inconsistent with the
string landscape picture.

If one assumes that for some reason numerically large cosmological constants get
forbidden, the landscape model leads to what we might call a landscape and forbiddenness
based MPP. In fact only a narrow band of values around zero cosmological constant would
then be allowed and the surviving vacua would obey MPP to the accuracy of the width
$w$ of this remaining band. However, this landscape and forbiddenness based MPP
would never be sufficiently accurate to become relevant for the main point of
the present article, according to which MPP ``transfers'' the cosmological
constant $\Lambda_{2}$ of the supersymmetric ``second vacuum''  to the physical
vacuum.

\section{Cosmological constant in Split SUSY scenario}
\label{cosmological}

In section 2 we argued that the supersymmetric Minkowski and physical vacua
can be degenerate in SUGRA models. We now try to
estimate the value of the cosmological constant in such models.
Thus we assume that the MPP is realized in Nature, leading to the formation
of a set of degenerate vacua with broken and unbroken local supersymmetry.
In the SUGRA models considered the low energy limit
of the theory in the vacuum with unbroken local supersymmetry is
described by a SUSY model in flat Minkowski space. So the
vacuum energy density of the corresponding supersymmetric states vanishes.
Because all vacua in the MPP inspired SUGRA models are
degenerate, the cosmological constant problem is thereby solved to first
approximation by assumption.

However the value of the cosmological constant may differ from zero in the
considered models. This occurs if non--perturbative effects in the ``observable
sector'' of the supersymmetric phase give rise to the breakdown of SUSY
at low energies. Then the MPP assumption implies that the physical
phase, in which local supersymmetry is broken in the hidden sector, has
the same energy density as a second phase, where supersymmetry
breakdown takes place in the observable sector.

The non-perturbative effects in the simplest SUSY models, like the minimal
supersymmetric standard model (MSSM), are extremely weak. Our strategy is to
estimate these effects in the vacuum with unbroken local supersymmetry
(the second vacuum) and thereby estimate the energy density in
this phase. This value of the cosmological constant
can then be interpreted as the physical value in our phase, by virtue of
the MPP.

If supersymmetry breaking takes place in the second vacuum, it is caused by the strong interactions. We assume the gauge couplings at high energies are 
identical in both vacua. Consequently the renormalization group running of 
the gauge couplings down to the scale
$M_S$, corresponding to the SUSY breaking scale in the physical vacuum,
are also the same in both vacua \footnote{ The gauge couplings
obey the renormalization group equations
$\ds\frac{d\log{\alpha_i(Q)}}{d\log{Q^2}}=\frac{b_i\alpha_i(Q)}{4\pi}$,
where $\alpha_i(Q)=g_i^2(Q)/(4\pi)$.}. Below the scale $M_S$ all
squarks and sleptons in the physical vacuum decouple and the corresponding beta functions
change. At the TeV scale, the beta functions in the physical vacuum change once again because of
the gradual decoupling of the gluino, neutralino and chargino. Using the value of
$\alpha^{(1)}_3(M_Z)\approx 0.1184\pm 0.0007$ and the matching condition
$\alpha^{(2)}_3(M_S)=\alpha^{(1)}_3(M_S)$, one finds the value of the strong coupling
in the second vacuum:
\be
\ds\frac{1}{\alpha^{(2)}_3(M_S)}=\ds\frac{1}{\alpha^{(1)}_3(M_Z)}-
\frac{\tilde{b}_3}{4\pi}\ln\frac{M^2_{g}}{M_Z^2}-\frac{b'_3}{4\pi}\ln\frac{M^2_{S}}{M_g^2}\, .
\label{22}
\ee
In Eq.(\ref{22}) $\alpha^{(1)}_3$ and $\alpha^{(2)}_3$ are the values of the strong gauge
couplings in the physical and second vacua, $M_g$ is the mass of the gluino, while $\tilde{b}_3=-7$
and $b'_3=-5$ are the one--loop beta functions of the strong gauge coupling in the SM and
Split SUSY scenario respectively.

The particles of the MSSM are, of course, all massless in the second supersymmetric phase,
where the EW symmetry is unbroken. So,
in the second vacuum, the renormalization group (RG) flow of the strong gauge coupling is determined
by the corresponding MSSM beta function, which exhibits asymptotically free behaviour ($b_3=-3$).
As a consequence $\alpha^{(2)}_3(Q)$ increases in the infrared region. Moreover the top quark
is massless in the second vacuum and the top quark Yukawa coupling grows in the infrared
with the increasing of $\alpha^{(2)}_3(Q)$. At the scale
\be
\Lambda_{SQCD}=M_{S}\exp\left[{\frac{2\pi}{b_3\alpha_3^{(2)}(M_{S})}}\right]\,,
\label{23}
\ee
where the supersymmetric QCD interactions become strong in the second vacuum, the
top quark Yukawa coupling is of the same order of magnitude as the strong gauge coupling.
The large Yukawa coupling of the top quark may result in the formation of a quark
condensate\footnote{Two of us have speculated \cite{nbs} that such a new phase might even
exist in the SM with a condensate of a 6 top and 6 antitop quark bound state.}
that breaks supersymmetry, inducing a non--zero positive value for the cosmological constant
\be
\Lambda \simeq \Lambda_{SQCD}^4\, .
\label{24}
\ee

In Fig.~1 the dependence of $\Lambda_{SQCD}$ for the second vacuum on the SUSY breaking
scale $M_S$ in the physical vacuum is examined.
In our numerical analysis we set $M_g=500\,\mbox{GeV}$. Because $b_3 > b'_3$ the
QCD gauge coupling below $M_S$ is larger in the physical vacuum than in the second vacuum.
Therefore the value of $\Lambda_{SQCD}$ is much lower than the QCD scale in the Standard
Model and diminishes with increasing $M_S$. When $M_S$ is of the order of 1 TeV, we obtain
$\Lambda_{SQCD}=10^{-26}M_{Pl} \simeq 100\,\mbox{eV}$. This leads to a vacuum energy density
($\Lambda\simeq 10^{-104} M_{Pl}^4$) which is much smaller than the electroweak scale
contribution in our vacuum $v^4 \simeq 10^{-62}M_{Pl}$. From Fig.~1 one can see that the
measured value of the cosmological constant is reproduced when $M_S\sim 10^{10}\,\mbox{GeV}$.
The value of the SUSY breaking scale, which leads to a reasonable agreement with the
observed vacuum energy density, depends on $\alpha_3(M_Z)$ and the gluino mass. As one can
see from Fig.~1, the dependence of $M_S$ on $\alpha_3(M_Z)$ is rather weak. With increasing
gluino mass the value of the SUSY breaking scale, which results in an appropriate value of the
cosmological constant, decreases. The results of our numerical analysis indicate that,
for $\alpha_3(M_Z)=0.116-0.121$ and $M_g=500-2500\,\mbox{GeV}$, the value of $M_S$ varies from
$2\cdot 10^9\,\mbox{GeV}$ up to $3\cdot 10^{10}\,\mbox{GeV}$.

The obtained prediction for the supersymmetry breaking scale can be tested. A striking
feature of the Split SUSY model is the extremely long lifetime of the gluino. In the
considered case, the gluino decays through a virtual squark to a quark antiquark pair
and a neutralino $\tilde{g} \rightarrow q\bar{q}+\chi_1^0$.
The large mass of the squarks then implies a long lifetime for the gluino. This lifetime
is given approximately by \cite{Dawson:1983fw}-\cite{Hewett:2004nw}
\be
\tau \sim 8\biggl(\ds\frac{M_S}{10^9\,\mbox{GeV}}\biggr)^4
\biggl(\ds\frac{1\,\mbox{TeV}}{M_g}\biggr)^5\,s.
\label{241}
\ee
When $M_S\gtrsim 10^{13}\,\mbox{GeV}$ the gluino lifetime becomes larger than the age of
the Universe. So long-lived gluinos would have been copiously produced during the very
early epochs of the Big Bang. They would have survived annihilation and would subsequently have
been confined in nuclear isotopes, which should be present in terrestrial matter \cite{43}.
Because the presence of such stable relics are ruled out by different experiments
(for $M_g\lesssim 10\,\mbox{TeV}$) \cite{42}, the supersymmetry breaking scale in the Split
SUSY scenario should not exceed $10^{13}\,\mbox{GeV}$ \cite{Giudice:2004tc}.
If, as is predicted, the SUSY breaking scale varies from $2\cdot 10^9\,\mbox{GeV}$
($M_g=2500\,\mbox{GeV}$) to $3\cdot 10^{10}\,\mbox{GeV}$ ($M_g=500\,\mbox{GeV}$),
then the gluino lifetime changes from $1\,\mbox{sec.}$ to $2\cdot 10^8\,\mbox{sec.}$
($1000\,\mbox{years}$). Thus the measurement of the gluino lifetime will allow an estimate
to be made of the value of $M_S$ in the Split SUSY model. The experimental signatures of long-lived
gluinos at colliders and in cosmic rays within the Split Supersymmetry scenario were explored
in \cite{Hewett:2004nw},\cite{split-susy-gluino}.

The observed value of the cosmological constant can also be reproduced for a much lower
supersymmetry breaking scale, if the MSSM particle content is supplemented by an
additional pair of $5+\bar{5}$ supermultiplets. The extra bosons and fermions would not affect
gauge coupling unification, because they form complete representations of $SU(5)$
(see for example \cite{29}). In the physical vacuum new bosonic states gain masses
around the supersymmetry breaking scale, while their fermionic partners survive to low
energies. In our numerical studies we assume that the masses of any new quarks are of the order
of the gluino mass, i.e. we set $M_q\simeq M_g\simeq 500\,\mbox{GeV}$. In the supersymmetric
phase the new bosons and fermions remain massless. The extra fermionic states do not much affect
the RG flow of gauge couplings in the Split SUSY scenario. For example, the one--loop beta
function that determines the running of the strong gauge coupling from the SUSY breaking
scale down to the TeV scale changes from $-5$ to $-13/3$. At the same time, the extra $5+\bar{5}$
supermultiplets give a considerable contribution to $b_3$ in the supersymmetric phase.
The corresponding one--loop beta function becomes $b_3=-2$. As a result $\alpha_3(Q)$ and
$\Lambda_{SQCD}$ decrease. In this case the observed value of the cosmological constant
can be reproduced even for $M_S\simeq 1\,\mbox{TeV}$ (see Fig.~1). However the Split
SUSY scenario has the advantage of avoiding the need for any new particles beyond those of
the MSSM, provided that $M_S\simeq 10^{10}\,\mbox{GeV}$.

\begin{figure}
\hspace*{-0cm}{\Large $\log[\Lambda_{SQCD}/M_{Pl}]$}\\[1mm]
\includegraphics[height=95mm,keepaspectratio=true]{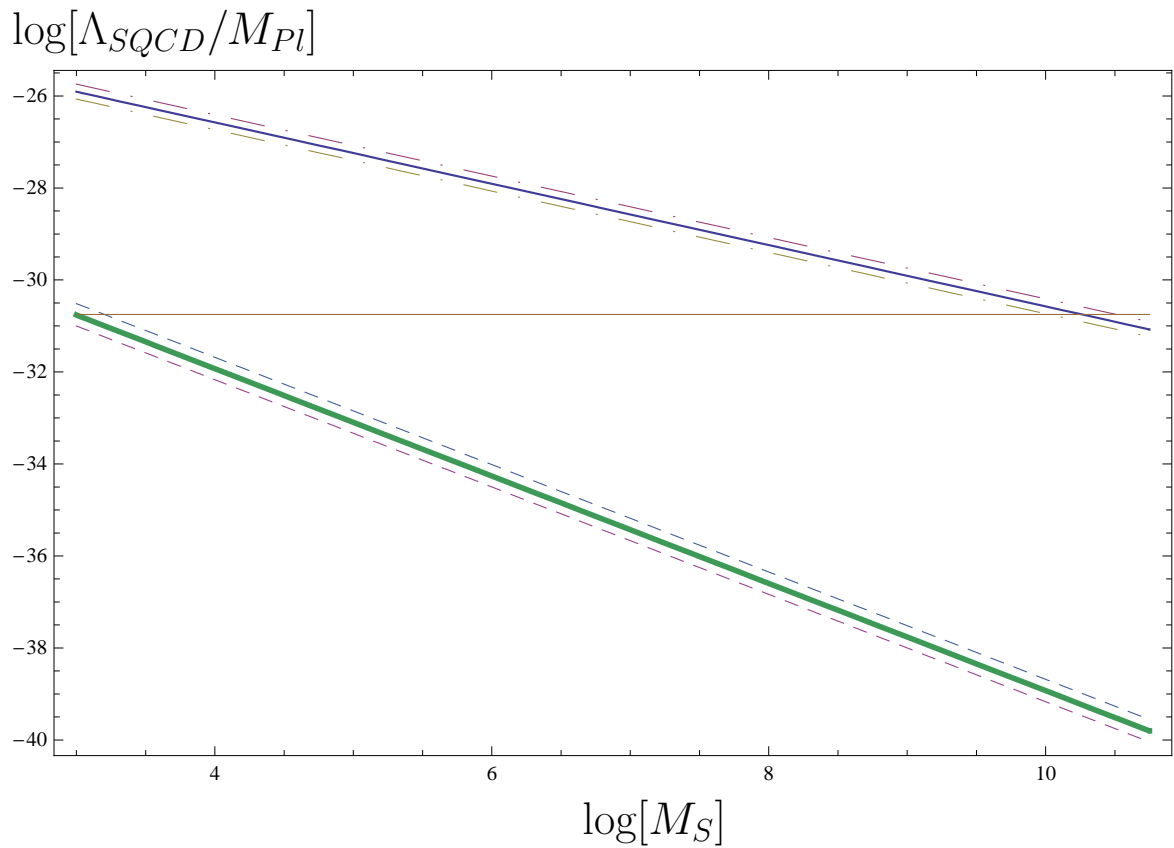}
\hspace*{7.5cm}{\Large $\log[M_S]$}\\[3mm]
\caption{The value of $\log\left[\Lambda_{SQCD}/M_{Pl}\right]$ versus $\log M_S$
for $M_q=M_g=500\,\mbox{GeV}$. The thin and thick solid lines correspond to the
Split SUSY scenarios with the pure MSSM particle content and the MSSM particle
content supplemented by an additional pair of $5+\bar{5}$ multiplets respectively.
The dashed and dash--dotted lines represent the uncertainty in $\alpha_3(M_Z)$.
The thin and thick solid lines are obtained for $\alpha_3(M_Z)=0.1184$,
the upper (lower) dashed and dash-dotted lines correspond to
$\alpha_3(M_Z)=0.116$ ($\alpha_3(M_Z)=0.121$). The horizontal line represents the
observed value of $\Lambda^{1/4}$. The SUSY breaking scale $M_S$ is measured in GeV.}
\label{essmfig1}
\end{figure}

\section{Conclusions}
\label{conclusion}

In the present article we have tried to estimate the value of the
cosmological constant in the Split Supersymmetry model, assuming
the existence of a set of degenerate vacua with broken and unbroken SUSY.
Using a sufficiently accurate principle of these vacua being degenerate
(MPP), a small cosmological constant is calculated in one of the vacua
and transferred to the other ones, especially to the physical one in
which we live. The idea is that we can calculate the cosmological
constant in the supersymmetric Minkowski (second) vacuum, having unbroken
local supersymmetry up to non-perturbative corrections coming from the
supersymmetric QCD scale in that vacuum. With reasonable Split SUSY
parameters - a SUSY breaking scale of $M_S \sim 10^{10}$ GeV - we obtain
the observed value of the cosmological constant.

We have argued that a set of such degenerate vacua can appear in $N=1$ supergravity
models. In general the presence of degenerate vacua in these models requires
an extra fine-tuning. Moreover an extra fine--tuning is normally needed to
ensure the existence of the vacuum in which the low--energy limit of the
considered theory is described by a pure supersymmetric model in flat
Minkowski space. The fine--tuning mentioned above can be alleviated within
the no--scale inspired SUGRA models in which the global symmetries,
that protect a zero value for the cosmological constant in all vacua and
preserve local supersymmetry, can be broken so that the MPP conditions are
fulfilled automatically without any extra fine-tuning at the tree--level.
In these models there can be a vacuum with softly broken supersymmetry
and spontaneously broken EW symmetry in the observable sector just like
in the physical vacuum in which we live. We demonstrated that the contributions
to the vacuum energy density, which originate from the breakdown of supersymmetry
and EW symmetry breaking in the physical vacuum, might get cancelled at the tree
level in these models.

However the inclusion of perturbative and non-perturbative corrections would
spoil the degeneracy of vacua. As a consequence an enormous fine tuning is
required, in general, to keep the vacuum energy densities in the vacua with
broken and unbroken supersymmetry to be the same. Nevertheless for our estimation
of the dark energy density we adopt such a fine-tuning assumption.
In other words we assume that there might be a mechanism (approximate global
symmetry, MPP or something else) in Nature that guarantees the
degeneracy of the vacua mentioned above. Moreover we restrict our consideration
to the simplest SUGRA models in which $f_a(T,\, z)\approx const$. In this limit
the gauginos are much lighter than squarks and sleptons, which is a
characteristic feature of the Split SUSY type spectrum.

The landscape of the metastable string theory vacua favors high-scale
SUSY breaking which could lead to Split SUSY. The enormous number of vacua
in String Theory can also lead to a large set of approximately degenerate
vacua, with broken and unbroken supersymmetry and tiny values of the vacuum
energy densities, if the vacua with large cosmological constants are forbidden.
Thus the Split SUSY scenario, landscape paradigm and the multiple point principle
might be different parts of the same picture. This landscape based MPP is
however not sufficiently accurate for the main calculation in this paper.

The main calculation in the present article concerns two of the several
vacua degenerate according to the MPP assumption. The first one is the
physical vacuum in which we live and in which we assume the physics is
described by the Split SUSY scenario. The second vacuum is one with unbroken
supersymmetry to first approximation. At high energy the physics is the same
in both these vacua, so that especially the running gauge couplings are equal
above the SUSY breaking scale $M_S$ in the physical vacuum. Also, in the second
vacuum the QCD ($SU(3)_C$) gauge coupling and the top quark Yukawa coupling
become large towards the infrared. At the scale $\Lambda_{SQCD}$ a top quark
condensate may get formed, giving rise to the breakdown of SUSY in
the supersymmetric phase and resulting in a non-zero and positive vacuum
energy density. The corresponding energy density is determined by the value
of the $SU(3)_C$ gauge coupling in the physical vacuum at high energies.
In order to calculate this energy density from the experimentally known
QCD coupling, the only unknown parameter is the SUSY breaking scale $M_S$
in the physical vacuum. For any value of
$M_S$ the calculated vacuum energy density in the supersymmetric phase
is rather small ($\lesssim 10^{-100}\,M_{Pl}^4$). The measured value of the
cosmological constant can be reproduced when $M_S\sim 10^{9}-10^{10}\,\mbox{GeV}$,
which leads to extremely long-lived gluinos. The gluino lifetime varies
from $1\,\mbox{sec.}$ to $2\cdot 10^8\,\mbox{sec.}$ This prediction can be
tested if such long-lived gluinos are observed in future experiments.
Previously \cite{Froggatt:2005nb} we showed that a reasonable value for
the dark energy density can be obtained even for $M_S\sim 1\,\mbox{TeV}$
if the MSSM particle content is supplemented by an extra pair of $5+\bar{5}$
supermultiplets. The scenario with extra $5+\bar{5}$ multiplets of matter and
the supersymmetry breaking scale in the $\mbox{TeV}$ range can also be tested
at the LHC in the near future. But with the Split SUSY scenario we avoid these
ad hoc $5+\bar{5}$ particles.

\section*{Acknowledgements}
R.N. would like to thank M.~Drees, K.~R.~Dienes, D.~Gorbunov, D.~R.~T.~Jones,
S.~F.~King, J.~P.~Kumar, S.~Pakvasa, D.~G.~Sutherland, X.~R.~Tata, B.~Thomas
and M.~I.~Vysotsky for fruitful discussions. The work of R.N. was supported
by the U.S. Department of Energy under Contract DE-FG02-04ER41291.
Also C.D.F. would like to acknowledge support from STFC in UK.


\newpage

\begin{thebibliography}{99}

\bibitem{1}
A.G.~Riess {\itshape et al.}, Astron.~J. {\bf 116}, 1009 (1998);
\newline
S.~Perlmutter {\itshape et al.}, Astrophys.~J. {\bf 517}, 565
(1999);
\newline
C.~Bennett {\itshape et al.}, Astrophys.~J.~Suppl. {\bf
148}, 1 (2003);
\newline
D.~Spergel {\itshape et al.}, Astrophys.~J.~Suppl.
{\bf 148}, 175 (2003).

\bibitem{2}
J.~Polonyi, Budapest preprint KFKI--1977 93 (1977).

\bibitem{10}
D.L.~Bennett and H.B.~Nielsen, Int.~J.~Mod.~Phys. A {\bf 9}, 5155
(1994);
\newline
D.L.~Bennett, C.D.~Froggatt and H.B.~Nielsen, in {\itshape
Proceedings of the 27th International Conference on High energy
Physics, Glasgow, Scotland, 1994}, p.~557; {\itshape Perspectives
in Particle Physics '94, World Scientific, 1995}, p.~255, ed. D.
~Klabu\u{c}ar, I.~Picek and D.~Tadi\'{c} [arXiv:hep-ph/9504294];
\newline
H.~B.~Nielsen and M.~Ninomiya, arXiv:hep-th/0701018;
\newline
H.~B.~Nielsen, Int.\ J.\ Mod.\ Phys.\  E {\bf 20}, 2049 (2011)
   [arXiv:1103.3812 [hep-ph]].

\bibitem{Froggatt:2003jm}
C.~Froggatt, L.~Laperashvili, R.~Nevzorov and H.~B.~Nielsen,
Phys.\ Atom.\ Nucl.\  {\bf 67} (2004) 582
[Yad.\ Fiz.\  {\bf 67} (2004) 601]
[arXiv:hep-ph/0310127].


\bibitem{Froggatt:2004gc}
C.~Froggatt, L.~Laperashvili, R.~Nevzorov and H.~B.~Nielsen,
in {\itshape Bled Workshops in Physics (ISSN: 1580-4992), Vol.~5,
No.~2, DFMA - zaloznistvo, Ljubljana, 2004}, p.~17
[arXiv:hep-ph/0411273].


\bibitem{Froggatt:2005nb}
C.~Froggatt, R.~Nevzorov and H.~B.~Nielsen,
Nucl.\ Phys.\  B {\bf 743} (2006) 133
[arXiv:hep-ph/0511259].


\bibitem{talks}
C.~D.~Froggatt, R.~Nevzorov and H.~B.~Nielsen,
J.\ Phys.\ Conf.\ Ser.\  {\bf 110} (2008) 072012
[arXiv:0708.2907 [hep-ph]];
C.~D.~Froggatt, R.~Nevzorov and H.~B.~Nielsen,
arXiv:0810.0524 [hep-th];
C.~D.~Froggatt, R.~Nevzorov and H.~B.~Nielsen,
AIP Conf.\ Proc.\  {\bf 1200} (2010) 1093
[arXiv:0909.4703 [hep-ph]].

\bibitem{4}
A.B.~Lahanas and D.V.~Nanopoulos, Phys.~Rep. {\bf 145} 1 (1987).

\bibitem{3}
H.P.~Nilles, Phys.~Rep. {\bf 110}, 1 (1984).

\bibitem{15}
J.~Ellis, M.K.~Gaillard, M.~G$\ddot{u}$naydin and B.~Zumino,
Nucl.~Phys. B {\bf 224}, 427 (1983).

\bibitem{12}
J.~Ellis, C.~Kounnas and D.V.~Nanopoulos, Nucl.~Phys. B {\bf 241},
406 (1984).

\bibitem{11}
E.~Cremmer, S.~Ferrara, C.~Kounnas and D.V.~Nanopoulos,
Phys.~Lett. B {\bf 133}, 61 (1983).

\bibitem{minimal-sugra}
R.~Barbieri, S.~Ferrara and C.~Savoy, Phys.~Lett. B {\bf 119}, 343
(1982); A.H.Chamseddine, R.Arnowitt, P.Nath, Phys.Rev.Lett. {\bf 49},
970 (1982); H.P.Nilles, M.Srednicki, D.Wyler, Phys.Lett. B
{\bf 120}, 345 (1983).

\bibitem{30}
G.F.~Giudice and A.~Masiero, Phys.~Lett. B {\bf 206}, 480 (1988);
\newline
J.A.~Casas and C.~Mu$\tilde{n}$oz, Phys.~Lett. B {\bf 306}, 288
(1993).

\bibitem{AMSB}
L.~Randall, R.~Sundrum,
Nucl.\ Phys.\ B  {\bf 557 } (1999)  79
[hep-th/9810155];
G.~F.~Giudice, M.~A.~Luty, H.~Murayama, R.~Rattazzi,
JHEP {\bf 9812 } (1998)  027
[hep-ph/9810442];
M.~Dine, N.~Seiberg,
JHEP {\bf 0703 } (2007) 040
[hep-th/0701023].



\bibitem{ArkaniHamed:2004fb}
N.~Arkani-Hamed, S.~Dimopoulos,
JHEP {\bf 0506} (2005) 073 [arXiv:hep-th/0405159].

\bibitem{Giudice:2004tc}
G.~F.~Giudice, A.~Romanino,
Nucl.\ Phys.\  B {\bf 699} (2004) 65
[Erratum-ibid.\  B {\bf 706} (2005) 65]
[arXiv:hep-ph/0406088].

\bibitem{ArkaniHamed:2004yi}
N.~Arkani-Hamed, S.~Dimopoulos, G.~F.~Giudice and A.~Romanino,
Nucl.\ Phys.\  B {\bf 709} (2005) 3
[arXiv:hep-ph/0409232].

\bibitem{dm-split-susy}
A.~Pierce,
Phys.\ Rev.\  D {\bf 70} (2004) 075006
[arXiv:hep-ph/0406144];
A.~Masiero, S.~Profumo and P.~Ullio,
Nucl.\ Phys.\  B {\bf 712} (2005) 86
[arXiv:hep-ph/0412058].

\bibitem{split-susy-spectrum}
I.~Antoniadis and S.~Dimopoulos,
Nucl.\ Phys.\  B {\bf 715} (2005) 120
[arXiv:hep-th/0411032];
B.~Kors and P.~Nath,
Nucl.\ Phys.\  B {\bf 711} (2005) 112
[arXiv:hep-th/0411201];
K.~S.~Babu, T.~Enkhbat and B.~Mukhopadhyaya,
Nucl.\ Phys.\  B {\bf 720} (2005) 47
[arXiv:hep-ph/0501079].

\bibitem{Murayama:2001ur}
H.~Murayama and A.~Pierce,
Phys.\ Rev.\  D {\bf 65} (2002) 055009
[arXiv:hep-ph/0108104].

\bibitem{Weinberg:1987dv}
S.~Weinberg,
Phys.\ Rev.\ Lett.\  {\bf 59} (1987) 2607.

\bibitem{anthropic-principle}
V.~Agrawal, S.~M.~Barr, J.~F.~Donoghue and D.~Seckel,
Phys.\ Rev.\  D {\bf 57} (1998) 5480 [arXiv:hep-ph/9707380];
C.~J.~Hogan,
Rev.\ Mod.\ Phys.\  {\bf 72} (2000) 1149 [arXiv:astro-ph/9909295];
M.~J.~Rees,
arXiv:astro-ph/0401424.

\bibitem{landscape}
S.~Kachru, R.~Kallosh, A.~D.~Linde and S.~P.~Trivedi,
Phys.\ Rev.\  D {\bf 68} (2003) 046005
[arXiv:hep-th/0301240];
L.~Susskind,
arXiv:hep-th/0302219;
T.~Banks, M.~Dine and E.~Gorbatov,
JHEP {\bf 0408} (2004) 058
[arXiv:hep-th/0309170];
M.~Dine, E.~Gorbatov, S.~D.~Thomas,
JHEP {\bf 0808} (2008) 098
[arXiv:hep-th/0407043].

\bibitem{Bousso:2000xa}
R.~Bousso and J.~Polchinski,
JHEP {\bf 0006} (2000) 006
[arXiv:hep-th/0004134].

\bibitem{large-number-of-vacua}
M.~R.~Douglas,
JHEP {\bf 0305} (2003) 046
[arXiv:hep-th/0303194];
S.~Ashok and M.~R.~Douglas,
JHEP {\bf 0401} (2004) 060
[arXiv:hep-th/0307049];
A.~Giryavets, S.~Kachru and P.~K.~Tripathy,
JHEP {\bf 0408} (2004) 002
[arXiv:hep-th/0404243];
J.~P.~Conlon and F.~Quevedo,
JHEP {\bf 0410} (2004) 039
[arXiv:hep-th/0409215].

\bibitem{high-energy-susy-breaking}
F.~Denef and M.~R.~Douglas,
JHEP {\bf 0405} (2004) 072
[arXiv:hep-th/0404116];
M.~R.~Douglas,
Comptes Rendus Physique {\bf 5} (2004) 965
[arXiv:hep-th/0409207].

\bibitem{Susskind:2004uv}
L.~Susskind,
arXiv:hep-th/0405189.

\bibitem{nbs}
C.~D.~Froggatt and H.~B.~Nielsen,
Surv. High Energy Phys. {\bf 18} (2003) 55 [arXiv:hep-ph/0308144];
C.~D.~Froggatt, L.~V.~Laperashvili and H.~B.~Nielsen,
Int. J. Mod. Phys. {\bf A20} (2005) 1268 [arXiv:hep-ph0406110];
C.~D.~Froggatt and H.~B.~Nielsen,
Phys. Rev. Lett. {\bf 95} (2005) 231301 [astro-ph/0508513];
C.~D.~Froggatt and H.~B.~Nielsen,
[arXiv:0810.0475 [hep-ph]];
C.~D.~Froggatt and H.~B.~Nielsen,
Phys. Rev. {\bf D80} (2009) 034033 [arXiv:0810.0475 [hep-ph]].

\bibitem{Dawson:1983fw}
S.~Dawson, E.~Eichten and C.~Quigg,
Phys.\ Rev.\  D {\bf 31} (1985) 1581.

\bibitem{Hewett:2004nw}
J.~L.~Hewett, B.~Lillie, M.~Masip and T.~G.~Rizzo,
JHEP {\bf 0409} (2004) 070
[arXiv:hep-ph/0408248].

\bibitem{43} S. Wolfram, Phys. Lett. B {\bf 82} (1979) 65;
C. B. Dover, T. K. Gaisser, G. Steigman, Phys. Rev. Lett.  {\bf 42} (1979) 1117.

\bibitem{42} J. Rich, M. Spiro, J. Lloyd--Owen, Phys. Rept.  {\bf 151} (1987) 239;
P. F. Smith, Contemp. Phys.  {\bf 29} (1988) 159; T. K. Hemmick et al. ,
Phys. Rev.  D {\bf 41} (1990) 2074.

\bibitem{split-susy-gluino}
A.~Arvanitaki, C.~Davis, P.~W.~Graham, A.~Pierce and J.~G.~Wacker,
Phys.\ Rev.\  D {\bf 72} (2005) 075011
[arXiv:hep-ph/0504210];
L.~Anchordoqui, H.~Goldberg and C.~Nunez,
Phys.\ Rev.\  D {\bf 71} (2005) 065014
[arXiv:hep-ph/0408284];
W.~Kilian, T.~Plehn, P.~Richardson and E.~Schmidt,
Eur.\ Phys.\ J.\  C {\bf 39} (2005) 229
[arXiv:hep-ph/0408088].

\bibitem{29}
R.~Hempfling, Phys.~Lett. B {\bf 351} (1995) 206.







\end{thebibliography}
\end{document}